\documentclass[twocolumn]{autart}    % Enable this line and disable the
\usepackage{cite,amssymb,amsmath,amsfonts}
\usepackage{color}
\usepackage{graphicx}          % Include this line if your
\usepackage{caption,subcaption}

\usepackage{tikz}
\usetikzlibrary{calc,shapes,arrows}

\usepackage{tikz}

\def\r{\mathbb R}

\begin{document}

\begin{frontmatter}
%\runtitle{Insert a suggested running title}  % Running title for regular
                                              % papers but only if the title
                                              % is over 5 words. Running title
                                              % is not shown in output.

%\title{Goodwin's oscillators with an additional negative feedback for modeling hormonal regulation systems \thanksref{footnoteinfo}}

\title{Local and global analysis of endocrine regulation as a
non-cyclic feedback system~\thanksref{footnoteinfo}}

\thanks[footnoteinfo]{This work was supported in part by
the NWO Domain TTW (project vidi-438730) and European Research Council (ERC-StG-307207).
The results have been partly reported at the IFAC Workshop on Periodic Control Systems PSYCO'16, Eindhoven, The Netherlands~\cite{taghvafard2016stability}.
Corresponding author H.~Taghvafard.}

\author[Gro]{Hadi Taghvafard}\ead{taghvafard@gmail.com},
\author[Delft,IPME,ITMO]{Anton V. Proskurnikov}\ead{anton.p.1982@ieee.org},
\author[Gro]{Ming Cao}\ead{m.cao@rug.nl}

\address[Gro]{Faculty of Science and Engineering, University of Groningen, %Nijenborgh 4, 9747 AG
Groningen, The Netherlands}
\address[Delft]{Delft Center for Systems and Control, Delft University of Technology, Delft, The Netherlands}
\address[IPME]{Institute for Problems of Mechanical Engineering of Russian Academy of Sciences (IPME RAS), St. Petersburg, Russia}
\address[ITMO]{Chair of Mathematical Physics and Information Theory, ITMO University, St. Petersburg, Russia}


\begin{abstract} % Abstract of not more than 200 words.
To understand the sophisticated control mechanisms of the human's endocrine system is a challenging task that is a crucial step towards precise medical treatment of many disfunctions and diseases. Although mathematical models describing the endocrine system as a whole are still elusive, recently some substantial progress has been made in analyzing theoretically its subsystems (or \emph{axes}) that regulate the production of specific hormones. Secretion of many vital hormones, responsible for growth, reproduction and metabolism, is orchestrated by feedback mechanisms that are similar in
structure to the model of simple genetic oscillator, described by B.C. Goodwin. Unlike the celebrated Goodwin model, the endocrinal regulation mechanisms are in fact known to have \emph{non-cyclic} structures and involve multiple feedbacks; a Goodwin-type model thus represents only a part of such a complicated mechanism. In this paper, we examine a non-cyclic feedback system of hormonal regulation, obtained from the classical Goodwin's oscillator by introducing an additional negative feedback. We establish global properties of this model and show, in particular, that
the \emph{local} instability of its unique equilibrium implies that almost all system's solution oscillate; furthermore, under additional restrictions these solutions converge to periodic or homoclinic orbits.
\end{abstract}

\end{frontmatter}

\section{Introduction}

\begin{figure}
\center
\includegraphics[height=4cm]{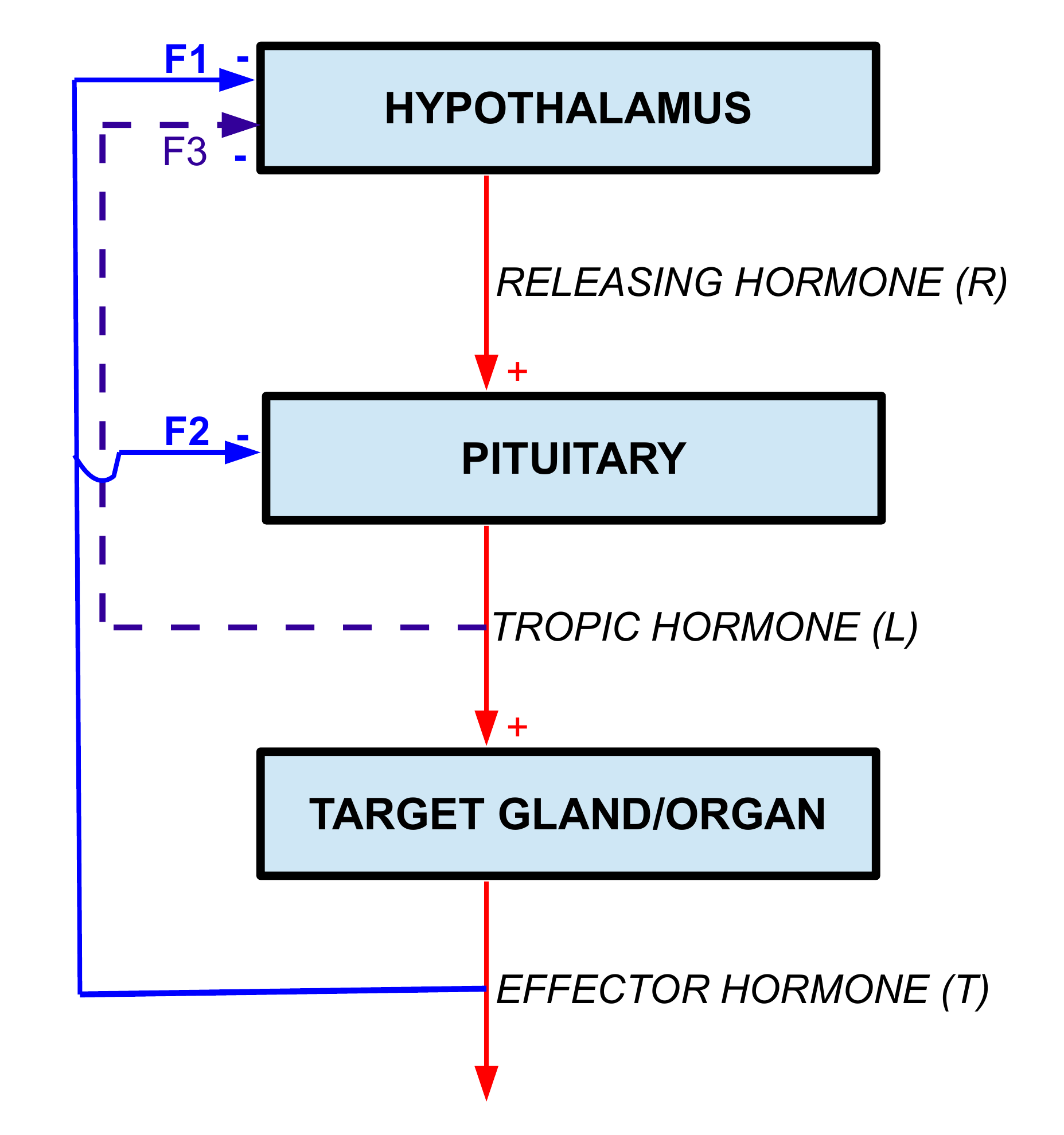}
\caption{The structure of a hypothalamic-pituitary axis~\cite{stear1975application}}
\label{fig.diagram}
\end{figure}

Hormones are signaling molecules that are secreted by glands and involved in many vital bodily functions. Sophisticated mechanisms of interactions between glands and hormones
couple them into the \emph{endocrine system}, whose mathematical modeling remains a challenging problem. At the same time, visible progress has been made in modeling
some of its subsystems, called \emph{axes}, and responsible for the secretion of specific hormones. Many processes in the body, including growth, metabolism, reproduction and stress resistance are controlled by the hypothalamic-pituitary (HP) neurohormonal axes. In the seminal work~\cite{stear1975application}
the feedback and feedforward \emph{control} mechanisms, lying the heart of the HP axes functioning, have been revealed; the first mathematical models had been proposed even earlier,
see e.g.~\cite{cronin1973} and references therein.
Regulatory centers in hypothalamus release special neurohormones, called \emph{releasing hormones} or \emph{releasing factors}~\cite{stear1975application}. Each of these hormones
stimulates the secretion of the corresponding \emph{tropic} hormone by the pituitary gland, which, in turn, stimulates some target gland or organ
to release the \emph{effector} hormone (Fig.~\ref{fig.diagram}). Besides its direct signaling functions, the effector hormone inhibits the production of the corresponding releasing and tropic hormones. These negative feedback loops maintain the concentrations of all three hormones within certain limits.

As many other biochemical systems, the endocrinal systems do not convergence to stable equilibria: the blood levels of hormones oscillate, exhibiting
both circadian (24-hour) and ultradian (short-period) oscillations~\cite{keenan-veldhuis2000,vinther2011minimal}. This oscillatory behavior resembles the dynamics of celebrated \emph{Goodwin's} model with three variables~\cite{Goodwin65}, considered as a ``prototypical biological oscillator''~\cite{Gonze:13Plos}.
Being originally proposed for a genetic intracellular oscillator~\cite{Goodwin65}, Goodwin-like models have been extensively used to describe the dynamics of HP axes, in particular,
the regulation of thyroid~\cite{cronin1973} and testosterone~\cite{smith1980hypothalamic} hormones. For Goodwin's oscillator and more general \emph{cyclic} feedback systems the profound mathematical results have been established, ensuring the existence of periodic orbits~\cite{Hastings77,HoriHara:11} in the case where the (only) system's equilibrium is unstable.
For the classical model from~\cite{Goodwin65} such an instability appears to be a restrictive condition, e.g. describing the feedback by the conventional Hill function~\cite{Gonze:13Plos}, the
corresponding Hill constant is required to be greater than $8$~\cite{smith1980hypothalamic,Thron91}. This restriction can be substantially relaxed
(yet not completely discarded~\cite{enciso2004stability}), taking into account the delays in hormone transporting~\cite{smith1983qualitative,das1994stability}. Other factors leading to oscillations are \emph{pulsatile} secretion of neurohormones~\cite{keenan1998biomathematical,churilov2009mathematical,churilov2014periodical} and stochastic noises~\cite{keenan1998biomathematical,keenan-veldhuis2000}.

\begin{figure}
\center
\includegraphics[height=4cm]{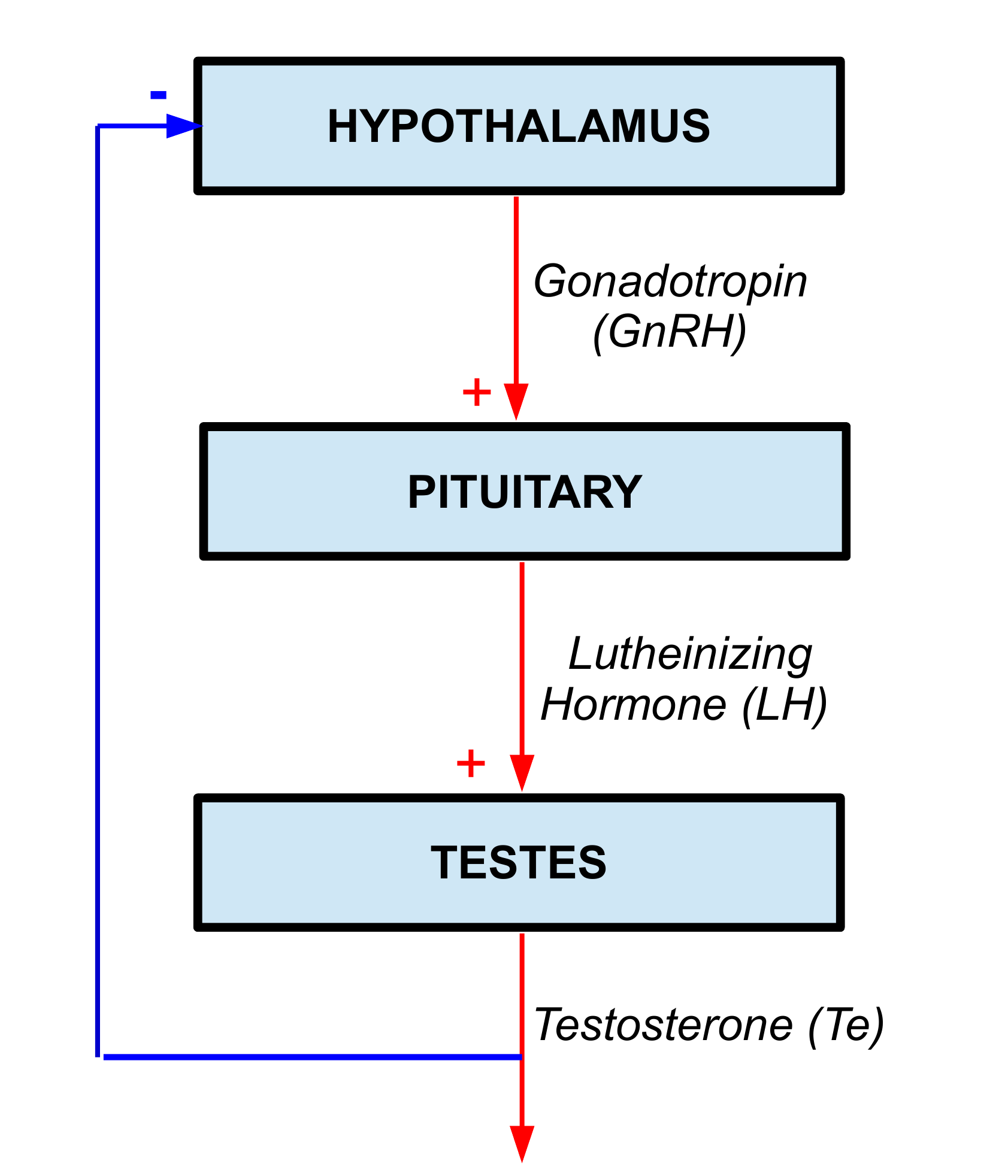}
\caption{The cyclic system of testosterone regulation~\cite{smith1980hypothalamic,churilov2009mathematical}}
\label{fig.diagram1}
\end{figure}

Being relatively well studied, Goodwin-type models however stipulate the presence of only one negative feedback loop from the effector hormone to the hypothalamus (F1 in Fig.~\ref{fig.diagram}).
This is illustrated by the models of testosterone regulation in males, examined in~\cite{smith1980hypothalamic,smith1983qualitative,das1994stability,churilov2009mathematical} and illustrated in Fig.~\ref{fig.diagram1}. At the same time, the complete mechanism of a HP axis involves multiple feedback loops~\cite{stear1975application}. The effector hormones inhibit the secretion of both releasing and tropic hormones, closing thus the \emph{long} negative feedback loops (F1,F2 in Fig.~\ref{fig.diagram}). Besides them, the \emph{short} feedback loop (F3) may also exist,
whose effect, however, is ignored by most of the existing mathematical models of endocrine regulation~\cite{bagatell1994direct,bing1991improved,keenan1998biomathematical,bairagi2008variability,Greenhalgh:09,vinther2011minimal,SriramDoyle2012}.
Being much weaker than the long feedbacks and ``most vulnerable''~\cite{stear1975application} among the three types of feedback mechanisms, the short feedback loops still lack experimental studies,
proving their ubiquity and revealing their role in endocrine regulation~\cite{stear1975application}.

Mathematical models, taking the existence of multiple feedback loops into account, have been recently proposed for the testosterone regulation in males~\cite{bing1991improved,Greenhalgh:09,Tanutpanit:15} and cortisol regulation~\cite{bairagi2008variability,vinther2011minimal,SriramDoyle2012}. 
Similar models with multiple feedback loops have been reported to describe the dynamics of some metabolic pathways~\cite{Sinha:87,Ghomsi:14}.
Unlike the classical Goodwin oscillators, these models do not have the cyclic structure, which makes the relevant results, ensuring the existence or absence of periodic solutions~\cite{Thron91,Hastings77,EncisoSmithSontag06,HoriHara:11,HoriTakadaHara:13} inapplicable; the mathematical analysis is limited to examination of local stability of equilibria points and the bifurcation analysis via Andronov-Hopf's techniques, ensuring the existence of periodic orbits only for \emph{some} values of the system's parameters.

In this paper, we examine a model of hormonal regulation with two negative feedbacks, which has been originally proposed in~\cite{bairagi2008variability} to describe the mechanism of cortisol regulation in the adrenal axis (hypothalamus-pituitary-adrenal cortex); our simulations (Section~\ref{sec.simul}) shows that it can also be applied
to testosterone regulation modeling. The model is similar in structure to the classical Goodwin oscillator, but involves two nonlinearities, standing for the negative feedbacks from the effector hormone to the releasing and tropic hormones (F1, F2 in Fig.~\ref{fig.diagram}). Unlike the original model in~\cite{bairagi2008variability}, we do not restrict these nonlinearities 
be identical, they also need not be Hill functions. To keep the analysis concise, in this paper we neglect the transport delays, discontinuities, describing the pulsatile secretion of neurohormones, and the effects stochastic noises. For the model in question, we develop the ``global'' theory, showing that its properties, in spite of the non-cyclic structure, are similar to ones of the Goodwin oscillator. In particular, under some assumptions, the \emph{local} unstability of the equilibrium implies the existence of periodic orbits and, furthermore, the convergence of \emph{almost any} solution to such an orbit. The latter statement, observed in simulations, have not been proved even for the classical Goodwin model.

This paper is organized as follows. Section~\ref{sec.model} introduces the model in question, whose local stability properties are examined in
Section~\ref{sec.local}. Section~\ref{sec.oscill} presents the main results of the paper, concerned with global properties of the system. 
Section~\ref{sec.simul} illustrates the model in question by numerical simulations. The results of the paper are proved in Section~\ref{sec.proof}. Section~7 concludes the paper.

%%%%%%%%%%%%%%%%%%%%%%%%%%%%%%%%%%%%%%%%%%%%%%%%%%%%%%%%%%%%%%%%%%%%%%%%%%%%%%%%%%
%%%%%%%%%%%%%%%%%%%%%%%%%%%%%%%%%%%%%%%%%%%%%%%%%%%%%%%%%%%%%%%%%%%%%%%%%%%%%%%%%%
\section{The Goodwin-Smith model and its extension}\label{sec.model}

We start with the conventional Goodwin's model~\cite{Goodwin65}, describing a self-regulating system of three chemicals, whose concentrations are denoted by $R$, $L$ and $T$ and evolve in accordance with the following equations
\begin{eqnarray}\label{SmithM}
	&&\dot{R} = -b_1 R + f(T),\nonumber\\
	&&\dot{L} = g_1 R - b_2 L,\\
	&&\dot{T} = g_2 L - b_3 T.\nonumber
\end{eqnarray}
The model~\eqref{SmithM} was originally used by B.C. Goodwin for modeling oscillations in a single self-repressing gene~\cite{Goodwin65}.
Our notation follows~\cite{smith1980hypothalamic}, where Goodwin's oscillator was proposed for modeling of the gonadal axis in male (Fig.~\ref{fig.diagram1}) and
$R,L,T$ stood, respectively, for the blood levels of the gonadotropin-releasing hormone (GnRH), lutheinizing hormone (LH) and testosterone (Te).

The constants $b_i>0$ (where $i=1,2,3$) stand for the clearing rates of the corresponding chemicals, whereas the constants $g_1,g_2>0$ and the \emph{decreasing} function $f:[0;\infty)\to (0;\infty)$ determine their production rates. Often $f(T)$ stands for the nonlinear \emph{Hill function}~\cite{Gonze:13Plos}
\begin{eqnarray}\label{Hill}
	f(T) = \frac{K}{1+\beta T^n}
\end{eqnarray}
where $K,\beta,n>0$ are constants. The releasing factor ($R$) drives the production of the tropic hormone ($L$), which in turn stimulates the secretion of the effector hormone ($T$); 
the positive constants $g_1,g_2$ stand for the corresponding \emph{feedforward} control gains. The effector hormone \emph{inhibits} the production of the releasing factor: since $f$ is a decreasing function, an increase in $T$ reduces the production rate $\dot R$, and vice versa. The nonlinearity $f(T)$ characterizes thus the negative \emph{feedback loop}.

In this paper, we consider a generalization of Goodwin's oscillator~\eqref{SmithM}, including \emph{two} negative feedbacks
\begin{eqnarray}\label{NewM}
	&&\dot{R} = -b_1 R + f_1(T),\nonumber\\
	&&\dot{L} = g_1 R - b_2 L + f_2(T),\\
	&&\dot{T} = g_2 L - b_3 T.\nonumber
\end{eqnarray}
A special case of~\eqref{NewM}, where $f_1,f_2$ stand for the Hill nonlinearities with the same Hill constants $n$, yet different gains $K_1,K_2$, has been proposed in~\cite{bairagi2008variability} 
to describe the dynamics of \emph{adrenal} axis: $R,L,T$ stand, respectively, for the levels of corticotropin-releasing
hormone (CRH), adrenocorticotropic hormone (ACTH) and cortisol. The nonlinearities $f_1,f_2$ describe the negative feedbacks F1,F2 in Fig.~\ref{fig.diagram}; the effect of short negative feedback (F3) is neglected. Unlike~\cite{bairagi2008variability}, in this paper we do not consider the effects of transport delays; at the same time, we substantially relax the assumptions imposed in~\cite{bairagi2008variability} on $f_1,f_2$. These nonlinear maps \emph{need not} be Hill functions, nor have the identical structure.
As discussed in the work~\cite{vinther2011minimal}, dealing with a similar model of cortisol regulation, the natural assumptions on these functions are their non-negativity
(which prevents the solutions from leaving the domain where $L,R,T\ge 0$), moreover, it is natural to assume that $f_1(T)>0$ since ``the feedbacks must not shut down hormone production completely''~\cite{vinther2011minimal}. Similar to the Goodwin model, two feedbacks are inhibitory, which implies that $f_1,f_2$ are non-increasing. We thus adopt the following assumption.
\begin{assum}\label{ass.basic}
The functions $f_1:[0;\infty)\to (0;\infty)$ and $f_2:[0;\infty)\to [0;\infty)$ are continuously differentiable and non-increasing, i.e.
$f_1'(T),f_2'(T)\le 0$ for any $T\ge 0$. The parameters $b_1,b_2,b_3,g_1,g_2>0$ are constant.
\end{assum}
Notice that we allow that $f_2(T)\equiv 0$; all of the results, obtained below, are thus applicable to the classical Goodwin oscillators~\eqref{SmithM}.
However, we are mainly interested in the case where $f_2\not\equiv 0$, which leads to the 
\emph{non-cyclic} structure of the system and makes it impossible to use mathematical tools developed for cyclic systems, such as criteria for global stability and periodic solutions existence~\cite{Thron91,HoriHara:11,Hastings77,HoriTakadaHara:13}. Unlike the existing works on multi-feedback models of hormonal regulation~\cite{bing1991improved,Greenhalgh:09,Tanutpanit:15,bairagi2008variability,vinther2011minimal,SriramDoyle2012}, our examination of the model~\eqref{NewM} is not limited
to establishing only local stability criteria and bifurcation analysis. In this paper, we are interested in the \emph{interplay} between local and global properties, revealed for the classical Goodwin's oscillator, namely, the existence of oscillatory solutions, provided that the (only) equilibrium of the system is unstable.

%%%%%%%%%%%%%%%%%%%%%%%%%%%%%%%%%%%%%%%%%%%%%%%%%%%%%%%%%%%%%%%%%%%%%%%%%%%%%
\section{Equilibria and local stability properties}\label{sec.local}

Since $R, L$ and $T$ stand for the chemical concentrations, one is interested in the solutions, starting in the positive octant $R(0),L(0),T(0)\ge 0$; this requires, due to Assumption~\ref{ass.basic},
that $R(t),L(t),T(t)>0$ for any $t>0$. Since $f_i(T)\le f_i(0)$, for all $T>0$, every solution is bounded. In particular, all the solutions are prolongable up to $\infty$.

%-------------------------------
The following properties of the system's equilibrium can be derived via straightforward computation, the details can be found in the conference paper~\cite{taghvafard2016stability}.
In view of Assumption \ref{ass.basic}, system \eqref{NewM} has a unique equilibrium point, namely $E^0=[R^0, L^0, T^0]$, where $R^0=\frac{1}{b_1}f_1(T^0), L^0=\frac{b_3}{g_2}T^0$, and $T^0>0$ is the unique root of
%\vspace{-0.2cm}
\begin{equation}\label{eq.t0}
\frac{b_1b_2b_3}{g_1g_2}T^0-\left[f_1(T^0)+\frac{b_1}{g_1}f_2(T^0)\right]=0.
\end{equation}
The coefficients of the characteristic equation corresponding to the Jacobian matrix of~\eqref{NewM} at $E^0$ are positive; it has a real negative root, and the two remaining roots are either complex-conjugated or real of the same sign~\cite{taghvafard2016stability}.
The Routh-Hurwitz criterion ensures the equilibrium is strictly stable (respectively, unstable) when $\Theta_0<0$ (respectively, $\Theta_0>0$) where we denote
\begin{equation}\label{H}
\begin{gathered}
%\begin{equation}\label{H}
	\Theta_0 \triangleq a_3 - a_1a_2 + g_2 \left[(b_2+b_3) f_2'(T^0) - g_1 f_1'(T^0)\right],\\
%\end{equation}
%with
%\begin{equation}\label{a1a2a3}
a_1 \triangleq b_1 + b_2 + b_3, a_2 \triangleq b_1b_2 + b_1b_3 + b_2b_3, a_3 \triangleq b_1 b_2 b_3.
%\end{equation}
\end{gathered}
\end{equation}
The aforementioned local stability properties of the equilibrium are summarized in the following lemma~\cite{taghvafard2016stability}.
\begin{lem}\label{lem.basic}
System~\eqref{NewM} has the unique equilibrium $E^0$ in the positive octant.
In both cases of $\Theta_0<0$ and $\Theta_0>0$, the equilibrium $E^0$ is hyperbolic.\footnote{For the dynamical system $\dot x=f(x)\in\r^m$ with $f(p)=0$, if the Jacobian $f'(p)$ has no eigenvalues with zero real parts, then $p$ is called \emph{hyperbolic}.}
If $\Theta_0=0$, then the two eigenvalues are complex-conjugated imaginary
numbers.
\end{lem}

In general, biochemical systems may have locally stable equilibria, whose existence does not exclude the possibility of periodic rhythms.
At the same time, for Goodwin's oscillator~\eqref{SmithM} the well-known ``secant condition''~\cite{Thron91}, being necessary and sufficient for local stability of the equilibrium, is in fact very close to the sufficient conditions of \emph{global} stability~\cite{ArcakSontag06}. In spite of some gap between the conditions of local and global stability, for Goodwin's models 
the equilibrium's instability is considered as the requirement of the biological feasibility; it is known, for instance, that Goodwin's oscillators and more general cyclic systems
with unstable equilibria have periodic orbits~\cite{HoriHara:11,Hastings77}. After the publication of the seminal Goodwin's paper~\cite{Goodwin65}, it was noticed~\cite{Griffith68,smith1980hypothalamic,Thron91} that for the Hill nonlinearity~\eqref{Hill} the equilibrium can be unstable (for some choice of the parameters $b_i,g_i$) if and only if $n>8$.
The following theorem extends the latter result to the generalized system~\eqref{NewM} and arbitrary decreasing functions $f_1(T),f_2(T)$.
We introduce an auxiliary function
\begin{equation}\label{eq.M}
M(T)\triangleq -Tf_1'(T)/f_1(T)>0, \quad \quad\forall T>0.
\end{equation}
\begin{thm}\label{thm.1}
Let the functions $f_1,f_2$ satisfy Assumption~\ref{ass.basic}. Then the following statements hold:
\begin{enumerate}
\item if $M(T)<8\,\forall T>0$ then $\Theta_0<0$ for any choice of $b_i,g_i>0$: the equilibrium of~\eqref{NewM} is stable;
\item if $M(T)\le 8\,\forall T>0$ then $\Theta_0\le 0$ for any $b_i,g_i>0$; the inequality is strict if $f_2(T)>0$ for any $T>0$;
\item if $M(T)>8$ for some $T>0$ then there exist parameters $b_i,g_i$ such that the equilibrium is unstable ($\Theta_0>0$) and, furthermore, the system has at least one non-constant
periodic solution.
\end{enumerate}
\end{thm}

Theorem~\ref{thm.1} will be proved in Section~\ref{sec.proof}. For the usual Goodwin-Smith model~\eqref{SmithM} it has been established in~\cite{smith1980hypothalamic}.
The existence of periodic solutions in statement 3) is based on the Hopf bifurcation theorem~\cite{Poore76}. However, the proof substantially differs from most of the existing results on the Hopf bifurcation analysis in \emph{delayed} biological oscillators~\cite{Greenhalgh:09, HuangCao:15, SunYuanCao:16}, proving the bifurcations at the ``critical'' delay values, under which the equilibrium loses its stability. To construct a one-parameter family of systems~\eqref{NewM}, satisfying the conditions of the Hopf bifurcation theorem, is not a trivial task (unlike the delayed case, where the delay is a natural parameter). One of such parameterizations has been proposed in~\cite{smith1980hypothalamic} for the model~\eqref{SmithM}, however, the complete and rigorous proof of the Hopf bifurcation existence has remained elusive.
\begin{rem}
While the necessary condition for instability is independent of the function $f_2(\cdot)$, the set of parameters $b_i,g_i$, for which the equilibrium is unstable, depends on it.
\end{rem}
\begin{rem}\label{rem.2}
Theorem~\ref{thm.1} does not imply that a periodic solution exists, \emph{whenever} the equilibrium in unstable.
The corresponding strong result holds for the Goodwin-Smith model~\eqref{SmithM} and more general cyclic systems~\cite{Hastings77,HoriHara:11}; in Section~\ref{sec.oscill} we extend this result
to a broad class of systems~\eqref{NewM}, where the nonlinearity $f_2(T)$ satisfies a special slope restriction, whose relaxation remains a non-trivial open problem. At the same time, as discussed in Section~\ref{sec.oscill}, the equilibrium's instability implies oscillatory behavior of the system~\eqref{NewM} in a some weaker sense.
\end{rem}
\begin{rem}\label{rem.3}
Although the conditions ensuring the equilibrium's global attractivity in the positive octant are close to the local stability~\cite{ArcakSontag06}, the
existence of (non-constant) periodic solutions in the case where $M(T)\le 8$ seems to be an open problem even for the Goodwin model~\eqref{SmithM}.
Furthermore, the Hopf bifurcation analysis in Section~\ref{sec.proof} shows that in the case where $M(T)>8$ there always exists a set of parameters $b_i,g_i$,
for which a periodic orbit coexists with the locally \textbf{stable} equilibrium.
\end{rem}

Applying Theorem~\ref{thm.1} to the case where $f_1(T)$ is the Hill function~\eqref{Hill}, one has $M(T)=-\frac{Tf_1'(T)}{f_1(T)}=n\frac{\beta T^n}{1+\beta T^n}$ and the condition $M(T)>8$
reduces to the well-known condition $n>8$. One arrives at the following. 
\begin{cor}\label{cor.1}
Suppose that $f_1(T)$ is the Hill function~\eqref{Hill} and $f_2$ satisfies Assumption~\ref{ass.basic}. Then the equilibrium of~\eqref{NewM} is stable whenever $n\le 8$. If $n>8$, then for some choice of $b_i,g_i>0$ the system has the unstable equilibrium and at least one periodic solution.
\end{cor}
The proof of Corollary~\ref{cor.1} is straightforward since the function $M(T)$, associated with the Hill function~\eqref{Hill}, is

It should be noticed that although the Hill functions~\eqref{Hill} with exponents $n>4$ are often considered to be non-realistic, Goodwin's models with $n>8$ adequately
describe some metabolic reactions (see~\cite{Gonze:13Plos} and references therein). More important, Goodwin-type oscillators with large Hill exponents $n$
naturally arise from \emph{model reduction} procedures~\cite{Gonze:13Plos}, approximating a long chain of chemical reactions by a lower-dimensional system.

\section{Oscillatory properties of the solutions}\label{sec.oscill}

As one can notice, Theorem~\ref{thm.1} does not establish any properties of system~\eqref{NewM} with some specific parameters $b_i,g_i$.
As discussed in Remark~\ref{rem.2}, it does not answer a natural question whether the equilibrium's instability $\Theta_0>0$ implies
any oscillatory properties of the system. In the case of the classical Goodwin-Smith system~\eqref{SmithM} ($f_2\equiv 0$) it is widely known that the local instability implies the existence of at least one periodic trajectory.
A general result from~\cite{Hastings77} establishes this for a general \emph{cyclic} system (with a sufficiently smooth right-hand side).
The cyclic structure of the system and the equilibrium's instability imply the existence of an \emph{invariant toroidal domain}~\cite{Hastings77}, and closed orbits in it correspond to
fixed points of the Poincar\'e map. This result, however, is not applicable to system~\eqref{NewM}. Another approach, used in~\cite{HoriHara:11,KimHoriHara:11,HoriTakadaHara:13} to examine
oscillations in gene-protein regulatory circuits, employs elegant results by Mallet-Parret~\cite{mallet1996poincare,MalletParret90}, extending the Poincar\'e-Bendixsson theory to Goodwin-type
systems. As discussed in Subsect.~\ref{subsec.mallet}, these results can be applied to system~\eqref{NewM} only if some additional restriction holds.

At the same time, when $\Theta_0>0$, one is able to prove an oscillatory property of the solutions, which was introduced by V.A. Yakubovich~\cite{Yakubovich73,TombergYak89} and states
that the solution is bounded, yet does not converge to an equilibrium. In the next subsection it is shown that, in fact, almost all solutions are oscillatory in this sense.

\subsection{Yakubovich-oscillatory solutions}\label{subsec.yakub}

Following~\cite{pogromsky1999diffusion}, we introduce the following definition.
\begin{defn}\label{scay}
	A scalar bounded function $\varrho:[0;\infty)\rightarrow\mathbb{R}$ is called Yakubovich-oscillatory, or \emph{$Y$-oscillation}, if
$\varliminf\limits_{t\to\infty}\varrho(t)<\varlimsup\limits_{t\to\infty}\varrho(t)$. A vector-valued function $x:[0;\infty)\rightarrow\mathbb{R}^m$ is called Y-oscillation if at least one of its elements $x_i(\cdot)$ is $Y$-oscillation. 
\end{defn}

In other words, $Y$-oscillation is a bounded function, having no limit as $t\to\infty$. Our next result shows that system~\eqref{NewM} with an unstable equilibrium has $Y$-oscillations; moreover, almost every solution is $Y$-oscillation. 
\begin{lem}\label{thm.2}
Suppose that system~\eqref{NewM} has an unstable equilibrium ($\Theta_0>0$). Then for any initial condition $(R(0),L(0),T(0))$, except for the points from some set of zero Lebesgue measure, the corresponding solution $(R(t),L(t),T(t))$ is Yakubovich-oscillatory as $t\rightarrow\infty$.
\end{lem}
Obviously, any periodic solution is Yakubovich-oscillatory, and the same holds for solutions converging to periodic orbits.
In general, a dynamical system can have other $Y$-oscillations, e.g. showing ``strange'' (chaotic) behavior. It is known, however, that solutions of the conventional Goodwin-Smith model~\eqref{SmithM} and many other
cyclic feedback systems~\cite{HoriHara:11,KimHoriHara:11,HoriTakadaHara:13} in fact exhibit a very regular behavior, similar to that of planar (two-dimensional) systems.
The corresponding elegant result has been established in the papers by Mallet-Parret~\cite{MalletParret90,mallet1996poincare}. A natural question, addressed in the next subsection, is
the applicability of the Mallet-Parret to the extended Goodwin-Smith model~\eqref{NewM}.

\subsection{The Mallet-Parret theorem for the extended Goodwin-Smith system: the structure of $\omega$-limit set}\label{subsec.mallet}

A point $x_*$ is said to be a \emph{limit point} (or a partial limit) of a function $x:[t_0;\infty)\to\r^m$ (where $t_0\in\r$) at $\infty$ if there exists a sequence $t_n\xrightarrow[n\to\infty]{} \infty$ such that
$x(t_n)\xrightarrow[n\to\infty]{} x_*$. The set of all limit points at $\infty$ is referred to as the \emph{$\omega$-limit} set of the function $x(\cdot)$ and denoted by $\omega(x)$.
In general, the $\omega$-limit set can be empty; however, for a \emph{bounded} function it is always non-empty, compact and connected~\cite{coddington1955theory}.
Similarly, for a function $x:(-\infty;t_0]\to\r^m$ one can define the limit points at $-\infty$, constituting the \emph{$\alpha$-limit set} of the function and denoted by $\alpha(x)$.
Obviously, the $\alpha$-limit set of a function $x(t)$ coincides with the $\omega$-limit of the function $\bar x(t)=x(-t)$.

The widely-known Poincar\'e-Bendixson theory for planar autonomous (time-invariant) systems states that the $\omega$-limit set of a bounded solution can be a closed orbit, an equilibrium point or
union of several equilibria and heteroclinic\footnote{Given a dynamical system $\dot x=f(x)\in\r^m$, its \emph{heteroclinic} solution is a globally defined non-constant solution $x(t):(-\infty;\infty)\to\r^m$, whose
limits at $\infty$ and $-\infty$ are equilibria. If these limits coincide, the solution is called \emph{homoclinic}.} trajectories, converging to them (it is possible that $\omega(x)$ is a union of an equilibrium and homoclinic trajectory, converging to it). Although this result is not applicable to the system of order $3$ or higher, it remains
valid for \emph{cyclic} systems~\cite{MalletParret90}, including the classical Goodwin oscillator~\eqref{SmithM} and similar models~\cite{HoriHara:11,HoriTakadaHara:13}.
In the more recent papers~\cite{elkhader92,mallet1996poincare,mallet1996systems} the Poincar\'e-Bendixson theory has been extended to tridiagonal systems (the result from~\cite{mallet1996poincare}
is applicable to even more general case of the delayed tridiagonal system). For the reader's convenience, we formulate the corresponding result below.

Consider the dynamical system of order $N+1$, where $N\ge 2$, described by the equations
\begin{equation}\label{general}
\begin{aligned}
	\dot{x}_0 &= h_0(x_0, x_1)\\
	\dot{x}_i &= h_i(x_{i-1}, x_i, x_{i+1}),\quad i=1,\ldots,N-1\\
	\dot{x}_N &= h_N(x_{N-1}, x_N, x_0),
\end{aligned}
\end{equation}
Here the functions $h_0(\xi,\zeta)$ and $h_i(\eta,\xi,\zeta)$, ($i=1,\ldots,N$), are $C^1$-smooth. It is assumed that all of them are \emph{strictly} monotone in $\zeta$; the functions
$h_i(\eta,\xi,\zeta)$ for $i=1,\ldots,N$ are also non-strictly monotone in $\eta$. That is, the $i$th chemical (where $i=1,\ldots,N$) influences the production rate of the $(i-1)$th one, positively
or negatively, and the $0$th chemical influences the production of the $N$th one. At the same time, chemical $i$ (where $i=0,\ldots,N-1$) may influence the production of chemical $(i+1)$; however,
such an influence is not necessary: it is allowed that $\frac{\partial h_{i+1}}{\partial x_i}\equiv 0$. 
The central assumption is that if the ``adjacent'' components influence each other, then the corresponding influences are
\emph{equally signed} (being either both stimulatory or inhibitory)
\begin{equation}\label{eq.compat}
	\frac{\partial h_{i+1}}{\partial x_i}\frac{\partial h_{i}}{\partial x_{i+1}}\ge 0\quad \forall i=0,\ldots,N-1.
\end{equation}
Applying a simple change of variables, one may assume, without loss of generality~\cite{mallet1996systems,mallet1996poincare} that
\begin{equation}\label{eq.compat1}
\begin{gathered}
\frac{\partial h_i(\eta,\xi,\zeta)}{\partial \eta}\ge 0,\,\delta_i\frac{\partial h_i(\eta,\xi,\zeta)}{\partial \zeta}>0,\\
\delta_i=
\begin{cases}
\phantom{\pm}1, i<N\\
\pm 1, i=N.
\end{cases}
\end{gathered}
\end{equation}
In this paper, we are interested in tridiagonal systems~\eqref{general} with a single equilibrium, for which the result of~\cite[Theorem 2.1]{mallet1996poincare} reduces\footnote{Formally,
the paper~\cite{mallet1996poincare} deals with delay systems, explicitly assuming that the delay is non-zero.
The results are, however, valid for tridiagonal systems~\eqref{general} without delays; as mentioned in~\cite[p.~442]{mallet1996poincare}, the corresponding result (under some additional restrictions) has been established in~\cite{elkhader92}.}
to the following simpler lemma.
\begin{lem}~\cite{HoriTakadaHara:13}\label{lem.mallet}
Let the $C^1$-smooth nonlinearities $h_i$ in~\eqref{general} satisfy the conditions~\eqref{eq.compat} and the system has only one equilibrium. Then the $\omega$-limit set of any \emph{bounded} solution can have one of the
following structural types: (a) closed orbit; (b) union of the equilibrium point and a homoclinic trajectory; (c) the equilibrium point (singleton).
\end{lem}

It should be noticed that Lemma~\ref{lem.mallet} cannot be \textit{directly} applied to system~\eqref{NewM} since the central assumption~\eqref{eq.compat} is violated: 
recall that the effector hormone's (T) production is driven by the tropic hormone (L) and, at the same time, inhibits its secretion (Fig.~\ref{fig.diagram}). It appears, however, that under an additional assumption, a one-to-one mapping $(R, L, T)\to (x_0, x_1, x_2)$ exists, transforming~\eqref{NewM} into the ``canonic'' form~\eqref{eq.compat1} with $N=3$ and $\delta_N=-1$. The corresponding extension is our main result.
\begin{thm}\label{thm.3}
Suppose that Assumption~\ref{ass.basic} holds and
\begin{equation}\label{eq.f2}
\sup_{T\ge 0}|f'_2(T)|\leq \frac{(b_3-b_2)^2}{4g_2}.
\end{equation}
Then any solution of~\eqref{NewM} has the $\omega$-limit set of one of the three types,
listed in Lemma~\ref{lem.mallet}. If the equilibrium is unstable, then almost any solution converges to either a periodic orbit or the closure of a homoclinic trajectory.
\end{thm}

It should be noticed that~\eqref{eq.f2} automatically holds for the classical Goodwin oscillator~\eqref{SmithM} (and, more generally, when $f_2$ is constant). 
Furthermore, if the equilibrium is unstable, the system~\eqref{SmithM}
has in fact no homoclinic orbits~\cite{HoriHara:11}. This leads to the following corollary.
\begin{cor}\label{cor.goodwin}
If the system~\eqref{SmithM} has unstable equilibrium, then it also has a (non-trivial) periodic orbit. Moreover, almost any solution converges to such an orbit.
\end{cor}

Whereas the first statement of Corollary~\ref{cor.goodwin} has been established for a very broad class of cyclic systems~\cite{Hastings77} and in fact does not rely on Mallet-Parret theory,
the second statement, confirmed numerical simulations, still has not been proved mathematically. 

For a general system~\eqref{NewM}, the inequality~\eqref{eq.f2} restricts the slope of the nonlinear function $f_2(\cdot)$. Numerical experiments, shown in Section~\ref{sec.simul}, show that
this condition is only sufficient, and the solutions' convergence to the periodic orbit may take place even if it is violated. In practice, the ``possibility of homoclinic trajectory is
negligibly small''~\cite{HoriTakadaHara:13} and they are usually not observed in the simulations. Such trajectories typically arise as ``limits'' of stable limit cycles, whose periods tend to infinity (this effect is called the ``homoclinic bifurcation''~\cite{Afraimovich:14}) and can be thus considered as ``degenerate'' periodic orbits. 

%---------------------------------------
\section{Numerical simulation}\label{sec.simul}
In this section, we give a numerical simulation, which allows to compare the behaviors of systems~\eqref{SmithM} and~\eqref{NewM}.
The model parameters  $b_1=0.1$ min$^{-1}$, $b_2=0.015$ min$^{-1}$, $b_3=0.023$ min$^{-1}$, $g_1=5$ min$^{-1}$ and $g_2=0.01$ min$^{-1}$ are chosen to comply with the existing experimental data reported in the works~\cite{cartwright1986model,das1994stability}, dealing with testosterone regulation (Fig~\ref{fig.diagram1}).

\begin{figure}[h]
	\centering
	\includegraphics[width=0.48\textwidth, height=6cm]{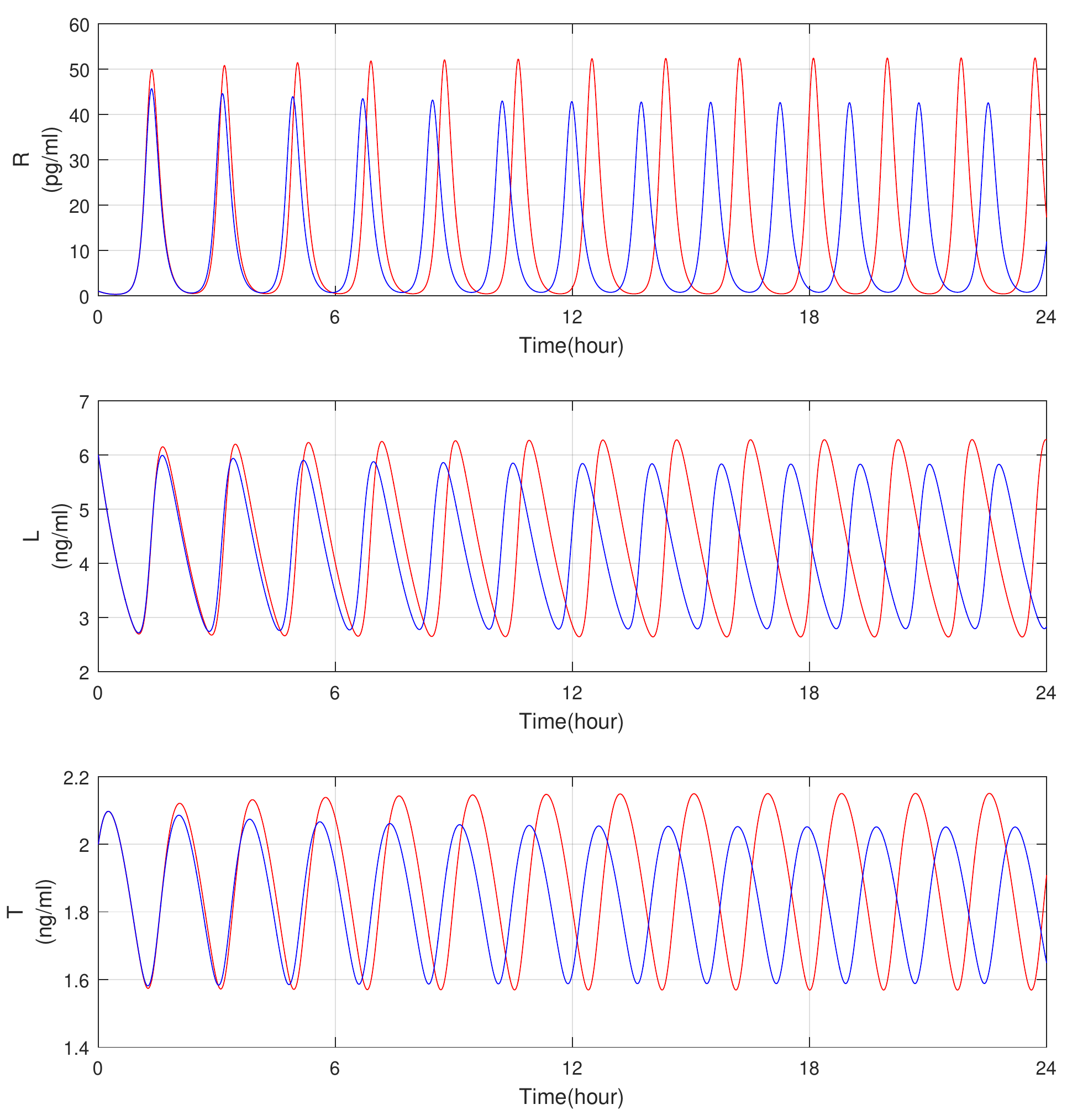}
	\caption{Red and blue plots show numerical simulations of systems~\eqref{SmithM} and~\eqref{NewM}, respectively, with the same initial conditions and parameter values.}\label{plot}
\end{figure}

The functions $f_1(T),f_2(T)$ were chosen of the Hill-type; as discussed in~\cite{vinther2011minimal,Gonze:13Plos} Hill's kinetics naturally arises in many biochemic and pharmacological systems.
Following~\cite{das1994stability}, the parameters of $f_1$ are considered to be $K_1=\beta_1=n=20$. To show the effect of the additional feedback $f_2$ on the oscillations of hormones, 
its parameters are chosen to be $K_2=m=20$ and $\beta_2=10$.
A straightforward calculation shows that the equilibria of systems~\eqref{SmithM} and~\eqref{NewM} are given by $E^{GS} = (0.0098, \ 3.2529, \ 1.4143)$ and  $E^{New} = (0.0094, \ 3.2589, \ 1.4169)$, respectively. Moreover, the quantity $\Theta_0$, defined in~\eqref{H}, for systems~\eqref{SmithM} and~\eqref{NewM} is given by $\Theta_0^{GS} = 1.5207\times 10^{-4}$ and $\Theta_0^{New} = 1.1590\times 10^{-4}$, confirming the instability of equilibria.
Both systems~\eqref{SmithM} and~\eqref{NewM} are plotted in Fig.~\ref{plot} for a time period of 24 hours with the same parameters and initial conditions $(R(0), L(0), T(0))=(1~pg/ml, ~6~ng/ml, ~2~ng/ml)$. Although nonlinearity $f_2$ considered in the example does not satisfy condition~\eqref{eq.f2}, system~\eqref{NewM} still have oscillatory behavior for parameters $b_i$ and $g_i$ considered above.

It is believed that exerting the feedback from Te to LH results in LH's amplitude reduction %~\cite{kandeel2007male},
and the effect of such a feedback is reduced by age~\cite{keenan2006ensemble,mulligan1997aging}.
As it is seen in Fig.~\ref{plot}, after some time, both amplitude and period of the oscillations of $R, L$ and $T$ in system~\eqref{NewM} become less than the corresponding ones in system~\eqref{SmithM}.
The amplitudes of oscillation for systems~\eqref{SmithM} and~\eqref{NewM}, calculated numerically, are given by $A^{GS}\approx(52~pg/ml,\  3.64~ng/ml,\  0.58~ng/ml)$, and $A^{New}\approx(41.75~pg/ml,\ 3.04~ng/ml,\  0.46~ng/ml)$, respectively.
Furthermore, the periods of oscillation for systems~\eqref{SmithM} and~\eqref{NewM} are given by $P^{GS}\approx 1.870$ and $P^{New}\approx 1.755$.
So the feedback $f_2(\cdot)$ influences both the amplitude and period of oscillations.
%%%%%%%%%%%%%%%%%%%%%%%%%%%%%%%%%%%%%%%%%%%%%%%%%%%%%%%%%%%%%%%%%%%%%%%%%5555

\section{Proofs of Theorems~\ref{thm.1} and~\ref{thm.3} and Lemma~\ref{thm.2}}\label{sec.proof}

We start with the proof of Theorem~\ref{thm.1}, extending the proofs from~\cite{Griffith68} and~\cite{smith1980hypothalamic}.
The proof employs the widely known McLaurin's inequality for the case of three variables
\begin{equation*}
	\frac{1}{3}(b_1+b_2+b_3)\geq \left({\frac{1}{3}\left(b_1b_2+b_1b_3+b_2b_3\right)}\right)^{\frac{1}{2}} \geq (b_1b_2b_3)^{\frac{1}{3}},
\end{equation*}
which holds for any $b_1,b_2,b_3>0$; both inequalities are strict unless $b_1=b_2=b_3$. It implies, in particular, that
\begin{equation}\label{eq.aux}
\frac{(b_1 + b_2 + b_3)(b_1b_2 + b_1b_3 + b_2b_3)}{b_1b_2b_3}\ge 9.
\end{equation}
Another result, used in the proof, is the Hopf bifurcation theorem~\cite{Poore76}.
This theorem deals with a one-parameter family of dynamical systems
\begin{equation}\label{eq.sys-mu}
\dot{x} = F(x,{\mu}),\quad \mu\in (-\varepsilon;\varepsilon).
\end{equation}
It is assumed that for $\mu=0$, the system has an equilibrium at $x_0$, for which $F(x,\mu)$ is $C^1$-smooth in the vicinity of $(x_0,0)$, and the Jacobian matrix $D_xF(x_0,0)$ has a pair of simple imaginary eigenvalues $\pm\imath\omega_0$ (where $\omega_0\ne 0$) and all other eigenvalues have non-zero real parts; in particular, $D_xF(x_0,0)$ is invertible. The implicit function theorem implies that for $\mu\approx 0$ there exists an equilibrium point $x(\mu)$ of system~\eqref{eq.sys-mu} (that is, $F(x(\mu),\mu)$), such that $x(0)=x_0$. The corresponding Jacobian $D_xF(x(\mu),\mu)$ has a pair of complex-conjugated eigenvalues
$\alpha(\mu)\pm\imath\omega(\mu)$, smooth for $\mu\approx 0$; here $\alpha(0)=0$ and $\omega(0)=\omega_0$.  The Hopf bifurcation theorem is as follows~\cite[Theorem 2.3]{Poore76}.
\begin{thm}\label{thm.hopf}
If $\alpha'(0)\ne 0$, the dynamical system~\eqref{eq.sys-mu} undergoes the Hopf bifurcation at $\mu=0$, that is, there exist $\varepsilon_0>0$ such that for
any $\mu\in (-\varepsilon_0,\varepsilon_0)\setminus\{0\}$ system~\eqref{eq.sys-mu} has a non-trivial periodic solution.
\end{thm}

\emph{Proof of Theorem~\ref{thm.1}}

Assuming that $(R^0,L^0,T^0)$ is an equilibrium of~\eqref{NewM} for some choice $b_i,g_i>0$ and applying~\eqref{eq.t0}, one obtains
	\begin{eqnarray}\label{g2}
		g_2=\frac{b_1b_2b_3 T^0}{g_1f_1(T^0)+b_1f_2(T^0)}.
	\end{eqnarray}
Substituting~\eqref{g2} into~\eqref{H} and dividing by $(b_1b_2b_3)$, the inequality~\eqref{eq.aux} and Assumption~\ref{ass.basic} imply the following
\begin{equation}\label{eq.theta0}
\begin{gathered}
\frac{\Theta_0}{b_1b_2b_3} =  \underbrace{\frac{T^0(b_2+b_3)f_2'(T^0)}{g_1f_1(T^0)+b_1f_2(T^0)}}_{\le 0}+\underbrace{\frac{g_1(-T^0f_1'(T^0))}{g_1f_1(T^0)+b_1f_2(T^0)}}_{\le M(T^0)}\\
- \frac{(b_1 + b_2 + b_3)(b_1b_2 + b_1b_3 + b_2b_3)}{b_1b_2b_3}+1 \le M(T^0)-8.
\end{gathered}
\end{equation}
The inequality~\eqref{eq.theta0} is strict unless $b_1=b_2=b_3$ and $f_2(T^0)=f_2'(T^0)=0$, implying thus statements~1 and~2.

We are now going to prove statement~3. Supposing that $M(T^0)>8$ for some $T^0>0$, let $R^0=\frac{1}{b_1}f_1(T^0)$ and $L^0=\frac{b_3}{g_2}T^0$. It can be easily noticed from~\eqref{eq.t0}
that any system~\eqref{NewM}, whose parameters satisfy the condition~\eqref{g2}, has the equilibrium at $(R^0,L^0,T^0)$. We are now going to design a one-parameter family of the systems~\eqref{NewM}
with this equilibrium, switching from stability to instability through a Hopf bifurcation.
To do this, we fix $b_1=b_2=b_3=b$ (where $b>0$ is chosen arbitrarily) and determine $g_2$ from~\eqref{g2}, leaving the parameter $g_1>0$ free.
It can be easily noticed from~\eqref{eq.theta0} that $\Theta_0=\Theta_0(g_1)$ is a smooth and strict increasing function of $g_1$, $\lim\limits_{g_1\to 0}\Theta_0(g_1)<0$ and $\lim\limits_{g_1\to\infty}\Theta_0(g_1)=M(T^0)-8>0$. Thus for sufficiently large $g_1>0$ the system has unstable equilibrium point. Furthermore, for $\varepsilon>0$ sufficiently small the image of $\Theta_0(\cdot)$
contains the interval $(-\varepsilon;\varepsilon)$; therefore, one can define the smooth inverse function $g_1=g_1(\mu)$ in such a way that $\Theta_0(g_1(\mu))=\mu$ for any $\mu=(-\varepsilon;\varepsilon)$.

We now claim that the one-parameter family of systems~\eqref{NewM} with $b_1=b_2=b_3=b>0$, $g_1=g_1(\mu)$ and $g_2=g_2(\mu)$ determined by~\eqref{g2} satisfies the conditions of Hopf bifurcation
theorem (Theorem~\ref{thm.hopf}). By definition, the Routh-Hurwitz discriminant~\eqref{H}, corresponding to a specific $\mu$, equals $\Theta_0(g_1(\mu))=\mu$; by Lemma~\ref{lem.basic} the system with
$\mu=0$ has a pair of pure imaginary eigenvalues. Considering the extension of these eigenvalues $\alpha(\mu)\pm\imath\omega(\mu)$ for $\mu\approx 0$, it is shown \cite[Appendix A]{taghvafard2016stability} that
\begin{equation}\label{eq.aux1}
\begin{gathered}
		2\alpha(\mu)\left[(a_1 + 2\alpha(\mu))^2 + (a_2 - g_2(\mu) f_2'(T^0))\right]=\mu
\end{gathered}
\end{equation}
(here $a_i$ are defined by~\eqref{H}). Differentiating~\eqref{eq.aux1} at $\mu=0$ and recalling that $\alpha(0)=0$, one arrives at $\alpha'(0)=\frac{1}{2\left[a_1^2+(a_2-g_2(0)f_2'(T^0)\right]}>0$.
Therefore, for $\mu\in (0;\varepsilon_0)$ (where $\varepsilon_0>0$) system~\eqref{NewM} with the aforementioned type has an unstable equilibrium at $(R^0,L^0,T^0)$ and at least one periodic solution. Notice however that for $\mu\in (-\varepsilon_0;0)$ the system also has a periodic solution in spite of the equilibrium's local stability (see Remark~\ref{rem.3}).\qed

\emph{Proof of Lemma~\ref{thm.2}}

Lemma~\ref{thm.2} is immediate from~\cite[Theorem~1]{pogromsky1999diffusion} since system~\eqref{NewM} (a) has the only equilibrium; (b) if $\Theta_0>0$ then
this equilibrium is \emph{hyperbolic} (there are no imaginary eigenvalues); (c) all solutions are uniformly ultimately bounded, that is, $C>0$ exists such that
\begin{equation*}\label{eq.aux3}
\varlimsup_{t\to\infty}(|R(t)|+|L(t)|+|T(t)|)\le C\;\forall R(0),L(0),T(0)>0.
\end{equation*}
The properties (a) and (b) follow from Lemma~\ref{lem.basic}; to prove (c) it suffices to notice that~\eqref{NewM} is decomposable as
$$
\dot X(t)=AX(t)+F(X(t)),\; X(t)=(R(t),L(t),T(t))^{\top},
$$
where $A$ is a Hurwitz matrix and $F(\cdot)$ is bounded.\qed

\emph{Proof of Theorem~\ref{thm.3}}

The restriction~\eqref{eq.f2} entails the existence of a one-to-one linear change of variables $(R,L,T)\mapsto (x_0,x_1,x_2)$, transforming~\eqref{NewM} into the general system~\eqref{general},
satisfying~\eqref{eq.compat1} with $\delta_N=-1$, $N=2$. Indeed, let $x_0 \triangleq T$, $x_1 \triangleq L+aT$ and $x_2 \triangleq R$,
where $a\in\r$ is a parameter to be specified later. The equations~\eqref{NewM} shape into~\eqref{general}, where
\begin{eqnarray*}
&&h_0(x_0,x_1) \triangleq g_2(x_1 - ax_0) - b_3x_0,\\
&&h_1(x_0,x_1,x_2) \triangleq \left(a(b_2-b_3)-a^2g_2\right)x_0 + f_2(x_0) + \\
&& \quad\quad\quad\quad\quad\quad\quad + g_1x_2 + (ag_2-b_2)x_1,\\
&&h_2(x_1,x_2,x_0) \triangleq -b_1 x_2 +f_1(x_0).
\end{eqnarray*}
Since $g_1,g_2>0$, the conditions~\eqref{eq.compat1} hold provided that $\frac{\partial h_1}{\partial x_0}\ge a(b_2-b_3)-g_2a^2-\sup|f_2'(T)|\ge 0$,
which always can be provided under the assumption~\eqref{eq.f2} by choosing appropriate $a\in\r$. Theorem~\ref{thm.3} now follows from Lemmas~\ref{thm.2} and~\ref{lem.mallet}: 
if the equilibrium is unstable, then almost all solutions do not converge.\qed
%%%%%%%%%%%%%%%%%%%%%%%%%%%%%%%%%%%%%%%%%%%%%%%%%%%%%%%%%%%%%%%%%%%%%%%%%%%
\section{Conclusions and future works}

A mathematical model for endocrine regulation has been examined, which extends the conventional Goodwin model by introducing the additional negative feedback.
We study the local properties of the extended model and their relations to \emph{global} properties, showing that the (locally) unstable equilibrium
implies that almost all solutions oscillate and (under some conditions) converge to periodic orbits. The results are based on the general criterion of oscillation
existence~\cite{TombergYak89} and the Mallet-Parret theory~\cite{mallet1996poincare}; they can be extended to many other models, e.g. the model from~\cite{vinther2011minimal}.
The relevant extensions are however beyond the scope of this manuscript due to the page limit.
Further extensions of the model, including transport delays and pulsatile feedback are the subject of ongoing research.

%%%%%%%%%%%%%%%%%%%%%%%%%%%%%%%%%%%%%%%%%%%%%%%%%%%%%%%%%%%%%%%%%%%%%%%%%%%%
\bibliographystyle{plain}
\bibliography{autosam}
\end{document}